\documentclass[fp,twocolumn]{jpsj3}
\usepackage{txfonts}
\usepackage{multirow}
\usepackage{url}
\usepackage{xcolor}

\title{Hybrid Optimization Method Using Simulated-Annealing-Based Ising Machine and Quantum Annealer}

\author{Shuta Kikuchi$^1$\thanks{skikuchi@keio.jp}, Nozomu Togawa$^2$, and Shu Tanaka$^{1,3,4,5}$}
\inst{$^1$Department of Applied Physics and Physico-Informatics, Keio University, Kanagawa 223-8522, Japan \\
$^2$Department of Computer Science and Communications Engineering, Waseda University, Tokyo 169-8555, Japan \\
$^3$Human Biology-Microbiome-Quantum Research Center (WPI-Bio2Q), Keio University, Tokyo 108-8345, Japan \\
$^4$Green Computing System Research Organization, Waseda University, Tokyo 162-0042, Japan \\
$^5$International Research Frontiers Initiative, Tokyo Institute of Technology, Tokyo, 108-0023, Japan} 

\abst{
Ising machines have the potential to realize fast and highly accurate solvers for combinatorial optimization problems. 
They are classified based on their internal algorithms. Examples include simulated-annealing-based Ising machines (non-quantum-type Ising machines) and quantum-annealing-based Ising machines (quantum annealers).
Herein we propose a hybrid optimization method, which utilizes the advantages of both types. 
In this hybrid optimization method, the preprocessing step is performed by solving the non-quantum-annealing Ising machine multiple times. 
Then sub-Ising models with a reduced size by spin fixing are solved using a quantum annealer.
The performance of the hybrid optimization method is evaluated via simulations using Simulated Annealing (SA) as a non-quantum-type Ising machine and D-Wave Advantage as a quantum annealer.
Additionally, we investigate the parameter dependence of the proposed hybrid optimization method.
The hybrid optimization method outperforms the preprocessing SA and the quantum annealing machine alone in fully connected random Ising models. 
}

\begin{document}
\maketitle

\section{Introduction}
\label{sec:introduction}

Ising machines have gained attention as accurate and efficient solvers for combinatorial optimization problems~\cite{Johnson2011, Yamaoka2016, Aramon2019, Inagaki2016, Goto2019, Maezawa2019, Motomura2020}. 
These problems find the optimal combination of decision variables to either minimize or maximize the objective function, while satisfying a set of constraints.
Combinatorial optimizations are applied in diverse fields such as machine learning~\cite{neven2009, Amin2015, Amin2018, Daniel2018}, materials science~\cite{Kitai2020,utimula2021quantum, inoue2022towards, endo2022phase, sampei2023quantum, tucs2023quantum, hatakeyama2023extracting}, portfolio optimization~\cite{rosenberg2016solving, Tanahashi2019}, protein folding~\cite{Perdomo-Ortiz2012}, traffic optimization~\cite{Neukart2017, irie2019quantum, Bao2021-a,Bao2021-b,Mukasa2021}, quantum compiler~\cite{naito2023isaaq}, and black-box optimization~\cite{Kitai2020, izawa2022continuous, seki2022black}.
To tackle combinatorial optimization problems using Ising machines, they employ a mapping technique to convert a problem into a mathematically constructed model in statistical mechanics called an \textit{Ising model} or its equivalent model called a \textit{quadratic unconstrained binary optimization (QUBO) model}.
QUBO is also referred to as the Unconstrained Binary Quadratic Programming problem (UBQP)~\cite{kochenberger2014unconstrained}.
Some combinatorial optimization problems can be formulated~\cite{lucas2014ising, Tanaka2017, Tanaka2020}.
By leveraging Ising or QUBO models, Ising machines search for the ground state, which represents the optimal solution to a problem.

An Ising model is defined on an undirected graph $G=(V, E)$, where $V$ and $E$ are sets of vertices and edges, respectively.
It consists of spins, interactions, and magnetic fields. 
The Hamiltonian (or energy function) $H$ of the Ising model is defined as
\begin{align}
  H = - \sum_{i\in V}h_{i}\sigma_{i} - \sum_{(i,j)\in E} J_{ij}\sigma_{i}\sigma_{j} ,
  \label{eq:H}
\end{align}
where $\sigma_i$ is the spin on the vertex $i \in V$ and takes a value of either $-1$ or $+1$.
The coefficients of the RHS in Eq.~\eqref{eq:H}, $h_{i}$ and $J_{ij}$, are the magnetic field on the vertex $i \in V$ and the interaction on the edge $(i, j) \in E$, respectively.

Various internal algorithms operate Ising machines~\cite{mohseni2022ising}.
This study focuses on quantum-annealing-based Ising machines (quantum annealers) and simulated-annealing-based Ising machines (non-quantum-type Ising machines). 
Quantum annealing (QA) is a heuristic algorithm that introduces quantum fluctuations.~\cite{kadowaki1998quantum}
Quantum annealers show quantum tunneling due to their quantum fluctuations.
Consequently, quantum annealers should be advantageous compared to classical methods.
Non-quantum-type Ising machines implement simulated annealing (SA), which is a popular algorithm to solve combinatorial optimization problems ~\cite{kirkpatrick1983optimization, johnson1989optimization, johnson1991optimization}.
The current advantage of non-quantum-type Ising machines over quantum annealers is that they can find lower-energy states of larger-scale Ising models.
Previous studies have compiled the specifications of various Ising machines~\cite{Oku2020,Kowalsky2021}. 

Several methods have been proposed to enhance the performance of Ising machines~\cite{rosenberg2016building, qbsolv, karimi2017effective, karimi2017boosting, irie2021hybrid, atobe2021hybrid, pelofske2021decomposition, pelofske2023solving}. 
These methods involve two steps.
First, smaller subproblems are generated using classical techniques or a quantum annealer through variable fixing or problem decomposition. 
Second, these subproblems are solved with an Ising machine.
Solving subproblems through this approach has produced improved metrics compared to directly solving the original problem using a quantum annealer in some studies~\cite{karimi2017boosting, pelofske2023solving}.
However, no studies have investigated hybrid methods that utilize different types of Ising machines.

In this study, we propose a hybrid optimization method that combines the advantages of a non-quantum-type Ising machine and a quantum annealer, examine the parameter dependence, and evaluate its performance via simulations.
The rest of this paper is organized as follows. 
Section~\ref{sec:hybrid_method}  introduces the hybrid method.
Section~\ref{sec:experimental_evaluations} performs the simulations using SA as a non-quantum-type Ising machine and D-Wave Advantage 4.1 (D-Wave Advantage) as a quantum annealing machine, with a fully connected Ising model that can be embedded in D-Wave Advantage.
The simulations investigate the influence of the parameters on the performance and evaluate the performance of the hybrid method compared to existing ones. 
Finally, Sect.~\ref{sec:conclusion} concludes our study and provides future research. 

\section{Hybrid optimization method}
\label{sec:hybrid_method}

\begin{figure*}[t]
    \includegraphics[clip,width=1.0\linewidth]{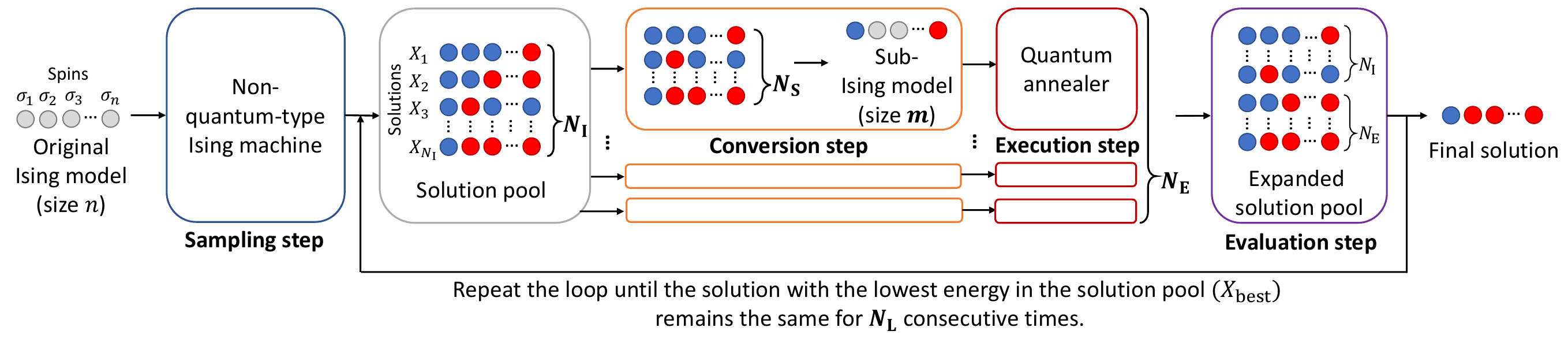}
    \caption{(Color online) Scheme of hybrid optimization method. The procedure consists of four steps: the sampling step, conversion step, execution step, and evaluation step. Gray, blue and red solid circles indicate that the spin has undetermined value, $-1$ and $+1$. Black arrows represent the flow. The sub-Ising model size ($m$), $N_{\rm I}$, $N_{\rm S}$, $N_{\rm E}$, and $N_{\rm L}$ are the parameters of the method.}
    \label{fig:hybrid_method_scheme}
\end{figure*}

Here we detail our hybrid optimization method inspired by the previous study~\cite{atobe2021hybrid}. 
Figure~\ref{fig:hybrid_method_scheme} depicts the proposed method.
The proposed method is composed of four steps. 

In the sampling step, $N_{\rm I}$ solutions are sampled using a non-quantum-type Ising machine and registered in the solution pool.
Because a non-quantum-type Ising machine shows a stochastic behavior, solutions with different spin states can be sampled.
The solution pool may contain duplicate solutions. 
This step ends by temporarily selecting the solution with the lowest energy in the solution pool as the best solution $X_{\textrm{best}}$.

In the conversion step, $N_{\rm S}$ solutions ($X_{1}$, $X_{2}$, ···, $X_{N_{\rm S}}$) are randomly selected from the solution pool, and using an approach called ``sample persistence''~\cite{karimi2017boosting,karimi2017effective}.
The spins are fixed to obtain a sub-Ising model.
Duplicate solutions may be included when performing $N_{\rm S}$ random selections.
Sample persistence assumes that spins with the same value across independently obtained samples are candidates to be fixed at a persistent value.
The remaining spins are a group of difficult spins to solve. 

Then sub-Ising models are obtained as follows.
Let $n$ and $m$ be the number of spins in the original Ising model and that in the sub-Ising model, respectively.
In the original Ising model, which is represented by Eq.~\eqref{eq:H}, let $S =$ \{ $\sigma_1$, $\sigma_2$, ..., $\sigma_n$ \} be a set of spins. 
First, $d_{i}$ is calculated for each spin of the $N_{\rm S}$ solutions to determine the spins not fixed at $-1$ or $+1$ into the sub-Ising model.
The value of $d_{i}$ is given as
\begin{align}
    d_{i}=\left|\sum^{N_{s}}_{k=1}\sigma_{i, k}\right|,
   \label{eq:d_i}
\end{align}
where $\sigma_{i, k}$ is the $i$-th spin in the $k$-th solution.
If $N_{\rm S}$ is even (odd), the minimum value of $d_{i}$ is $0$ ($1$).
The minimum value of $d_{i}$ is the most unstable spin, whereas the maximum value of $d_{i}$ is $N_{\rm S}$, which is a persistent spin.
To obtain the more unstable spins, collect $m$ spins of the sub-Ising model in order of the smallest $d_{i}$.
Let $S'$ be a set of spins in the sub-Ising model.

Next, randomly select a tentative solution from the $N_{\rm S}$ solutions.
Let $\tilde{\sigma}_i =$ ($\tilde{\sigma}_1$, $\tilde{\sigma}_2$, ··· , $\tilde{\sigma}_n$) be a spin state of the tentative solution, where $\tilde{\sigma}_i$ is $-1$ or $1$.
Then, the Hamiltonian of the sub-Ising model is generated from $S'$ and $\tilde{\sigma}_i$ as
\begin{align}
  H_\mathrm{sub-Ising \ model} = - \sum_{\substack{i \\ \sigma_{i}\in S'}}L_{i}\sigma_{i} - \sum_{\substack{(i,j) \\ \sigma_{i},\sigma_{j}\in S'}} J_{ij}\sigma_{i}\sigma_{j} + C ,
  \label{eq:H_sub}
\end{align}
where
\begin{align}
  L_{i} = h_{i} + \sum_{\substack{j \\ \tilde{\sigma}_{j}\notin S'}} J_{ij}\tilde{\sigma_{j}},
  \label{eq:L_i}
\end{align}
\begin{align}
  C = - \sum_{\substack{i \\ \tilde{\sigma_{i}}\notin S'}} h_{i} \tilde{\sigma}_{i} - \sum_{\substack{(i,j) \\ \tilde{\sigma}_{i},\tilde{\sigma}_{j}\notin S'}} J_{ij}\tilde{\sigma}_{i}\tilde{\sigma}_{j}.
  \label{eq:C}
\end{align}

In the execution step, a quantum annealer searches for lower-energy states of the sub-Ising model.
The spins contained in $S'$ of the tentative solution are updated with the spin state of the sub-Ising model.
This becomes the new solution.
Then, the conversion step and the execution step are repeated $N_{\rm E}$ times to obtain $N_{\rm E}$ new solutions are obtained.
These new solutions are added to the solution pool, constructing an expanded solution pool.

In the evaluation step, $N_{\rm I}$ low energy solutions are collected from the expanded solution pool to create a new solution pool. 
The lowest energy solution in the new solution pool is updated as the new $X_{\textrm{best}}$.
Then the flow from the solution pool creation to the evaluation step is iterated using the new solution pool.
This is repeated until $X_{\textrm{best}}$ remains the same for $N_{\rm L}$ consecutive times.
Finally, the hybrid optimization method outputs the lowest energy in the solution pool $X_{\textrm{best}}$ at the end of the iteration as the solution.

\section{Experimental evaluations}
\label{sec:experimental_evaluations}

\subsection{Setup of experiment}
\label{subsec:setup}

To demonstrate the effectiveness of our hybrid optimization method, we considered a fully connected random Ising model, which was represented as a complete graph.
The choice of an Ising model was motivated by the fact that all spins show a uniform edge density in all sub-Ising models once the spin is fixed on a complete graph. 
Using other Ising models may lead to various graph structures or the generation of isolated spins in each sub-Ising model, making it difficult to consistently evaluate the hybrid method due to variations in the hardware’s influence on D-Wave Advantage. 
To mitigate this, we adopted a complete graph. 

In addition, when the magnetic field coefficients $h_{i}=0$, Ising models display a twofold degeneracy for all states, rendering a sub-Ising model generation of the proposed hybrid method ineffective. 
As a countermeasure, we set one randomly chosen spin to either $-1$ or $+1$ as described in the previous study~\cite{karimi2017boosting, karimi2017effective}.
However, in this study, we set the magnetic field coefficient $h_{i}\neq0$ and assumed no trivial degeneracy.
The coefficients of the magnetic field and the interactions were randomly selected according to a Gaussian distribution with a mean of zero and a standard deviation of unity.
It should be noted that zero was excluded for both the interactions and the magnetic fields. 

In this study, we performed simulations using SA as a non-quantum-type Ising machine and D-Wave Advantage as a quantum annealer.
Most non-quantum-type Ising machines employ SA as the basis of its internal algorithms~\cite{Yamaoka2016, Yoshimura2017, Okuyama2017, Aramon2019 ,yamamoto2020, FixAE}.
The SA preprocessing parameters were a randomly set initial state and an initial temperature $T_\textrm{initial}$ of $\lceil 2v_\mathrm{max} \rceil$. 
Here, $v_\textrm{max}$ waas the maximum value among the $v_{i}$ defined by $v_{i}=\left|h_{i}+\sum_{j}{J_{ij}}\right|$, which is the absolute value of the sum of the magnetic fields and the interactions for $i$-th spin in the original Ising model.
The initial temperature was set sufficiently high to accommodate for the transition between arbitrary states in the beginning of SA.
The temperature schedule was set to the power-law decay for every outer loop, which is given by $T(t)=T_\textrm{initial} \times r^t$, where $r$ is the cooling rate and $t$ is the $t$-th outer loop.
The outer loop was set for each original Ising model.
The cooling rate was set such that the final temperature was equal to 0.1, which is sufficiently low on the energy scale of the original Ising model.
The inner loop was set at 160, which is the size of the original Ising model.

When using D-Wave Advantage, one original Ising model was annealed 100 times, and the best solution was selected.
The other parameters of D-Wave Advantage were set to their default values~\cite{Ocean}.
To compare the performance of D-Wave Advantage, the number of spins in the original Ising model was set to the maximum number that can be embedded in D-Wave Advantage.
D-Wave Advantage used in this study had $5,627$ qubits and a Pegasus graph topology~\cite{dattani2019pegasus}, allowing a complete graph of $177$ spins to be embedded using miner embedding~\cite{DW_update}.
However, several qubits could not be used due to defects.
After conducting multiple tests, a stable embeddable size of $160$ spins was set as the original Ising model size.

\subsection{Numerical experiment}
\label{subsec:numerical_experiment}

\begin{figure}
    \includegraphics[clip,width=1.0\linewidth]{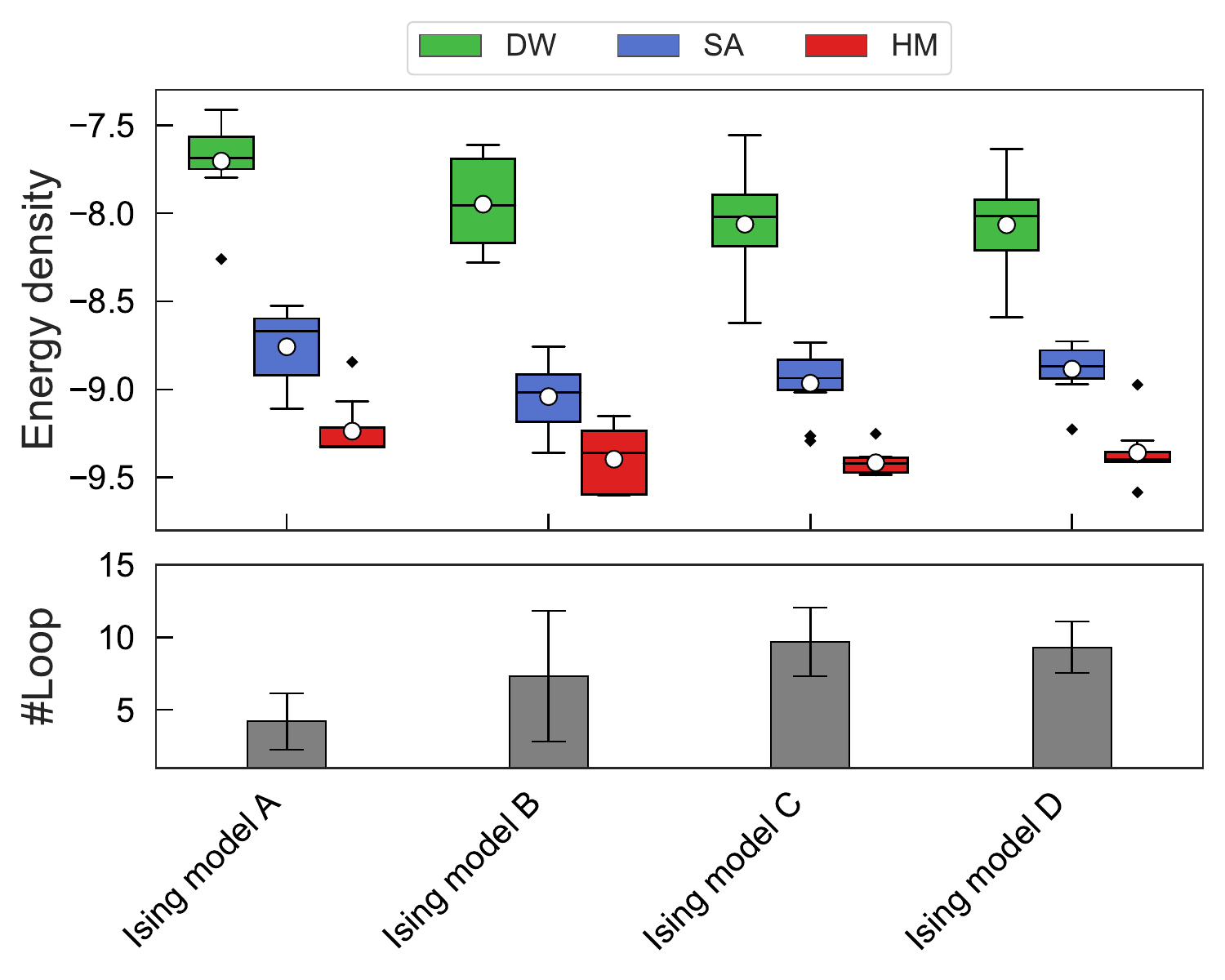}
    \caption{(Color online) (Top) Energy densities by different solvers in multiple original Ising models. White circles are the average of ten runs. Top and bottom bars denote the highest and lowest energy densities, respectively. Black diamonds denote the outliers. (Bottom) Number of loops by the hybrid method. Error bars are the standard deviations obtained in ten runs.}
    \label{fig:comparison_Ising_model}
\end{figure}

First, we performed the proposed hybrid optimization method using several original Ising models to evaluate its performance.
Table~\ref{table:parameters_1} summarizes the parameters for the outer loop, sub-Ising model size ($m$), $N_{\rm I}$, $N_{\rm S}$, $N_{\rm E}$, and $N_{\rm L}$.
D-Wave Advantage and preprocessing SA were considered for the reference solution accuracy.
Figure~\ref{fig:comparison_Ising_model} shows the energy densities of the solutions obtained by D-Wave Advantage (DW), preprocessing SA (SA), and hybrid optimization method (HM).
The SA data use $X_{\textrm{best}}$ obtained from each solution pool as the preprocessing, and the energy density is the internal energy per spin (i.e., $H/n$).
Figure~\ref{fig:comparison_Ising_model} also shows the number of loops spent until the completion of the hybrid method is complete.
The number of loops represents the count of returning to the solution pool creation from the execution step. 
As an example, when $N_{\rm L} = 3$ and $X_{\textrm{best}}$ was not updated at all, the number of loops is $2$.
These data were obtained from the average and standard deviation for ten simulations.
The hybrid method achieved a higher solution accuracy than either of DW or SA.

\begin{table}
  \centering
  \caption{Parameters of SA and hybrid optimization methods. $N_{\rm I}$ is the number of solution samples to be obtained at the sampling step, $N_{\rm S}$ is the number of reference original Ising models for generating sub-Ising models at the conversion step, $N_{\rm E}$ is the number of newly obtained solutions through repetition of the conversion and execution steps, and $N_{\rm L}$ is the number of repeats of the same $X_{\textrm{best}}$ to end the hybrid method. These parameters are described in Sect.~\ref{sec:hybrid_method} and Fig.~\ref{fig:hybrid_method_scheme}.}
  \label{table:parameters_1}
  \begin{tabular}{ccc} 
  \hline
    Optimization method & Parameter & Value \\
    \hline
    SA & Outer loop & 10  \\
    \hline
    \multirow{5}{*}{\shortstack[c]{Hybrid optimization \\ method}}
    & \shortstack[c]{Sub-Ising model \\ size ($m$)} & \raisebox{0.5em}{80} \\
    & $N_{\rm I}$ & 20  \\
    & $N_{\rm S}$ & 10  \\
    & $N_{\rm E}$ & 20  \\
    & $N_{\rm L}$ & 3   \\
    \hline
  \end{tabular}
\end{table}

\begin{figure}
    \begin{center}
    \includegraphics[clip,width=0.9\linewidth]{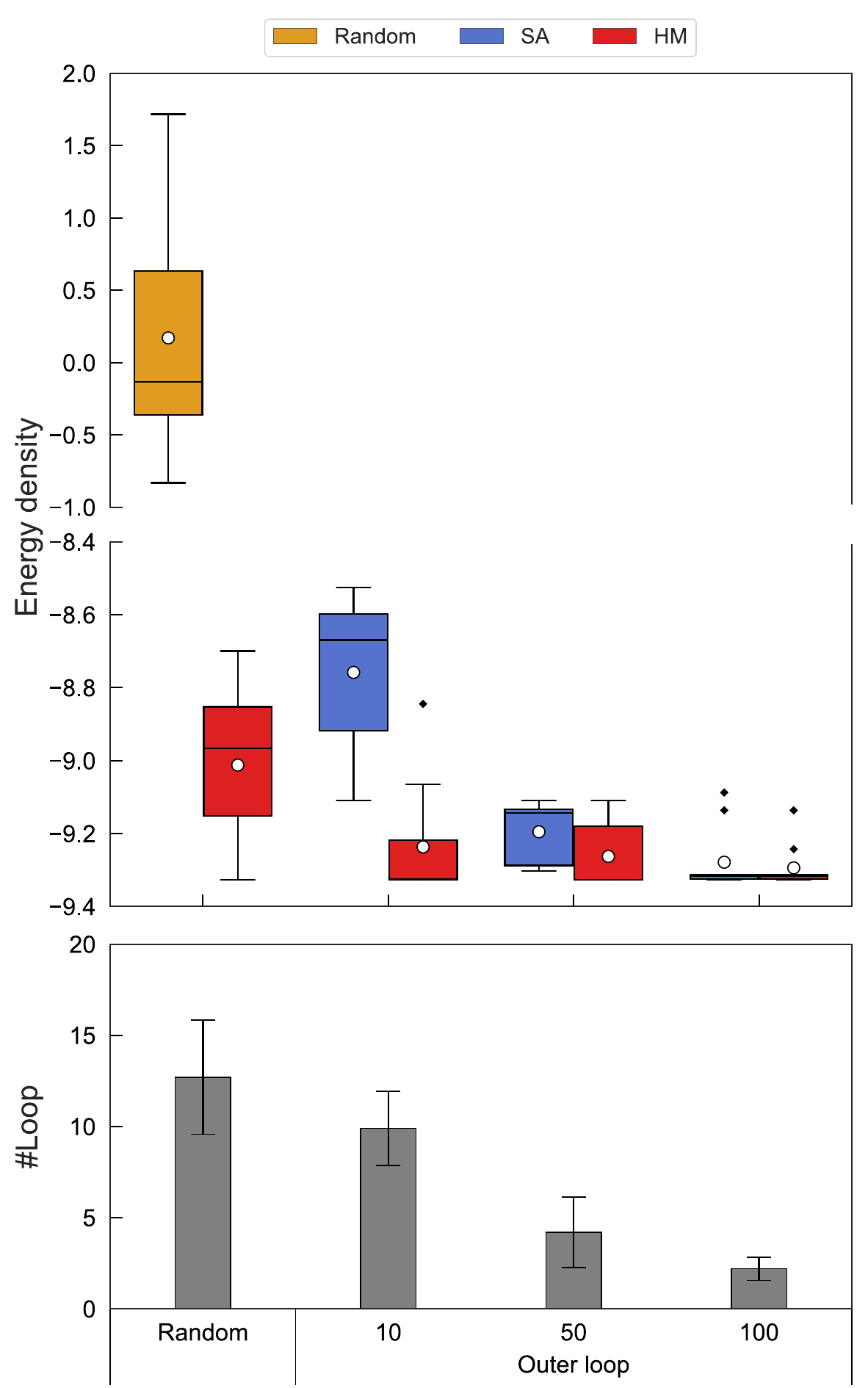}
    \caption{(Color online) Performance of hybrid optimization methods in preprocessing with different solution accuracies. (Top) Energy densities of preprocessing and the hybrid optimization method. White circles are the average of ten runs. Top and bottom bars denote the highest and lowest energy densities, respectively. Black diamonds denote outliers. Scales of energy densities differ between the upper and lower across the omitted portions. (Bottom) Number of loops by the hybrid optimization method. Error bars are standard deviations obtained in ten runs.}
    \label{fig:comparison_preprocessing}
    \end{center}
\end{figure}

Second, we evaluated the solution accuracy of the preprocessing dependency.
The hybrid method was performed using four different preprocessings: random spin states (Random), performing SA with the outer loop set to 10, 50, or 100.
The parameters of the hybrid method are the same as those in Table~\ref{table:parameters_1}.
Figure~\ref{fig:comparison_preprocessing} shows the energy densities and the number of loops with the different preprocessings.
Regardless of the solution accuracy of the preprocessing, the hybrid method improved the solutions from that of preprocessing. 
Moreover, even for the preprocessing with an extremely low solution accuracy (e.g., Random), the hybrid method gave a certain level of solution accuracy.
However, a lower preprocessing solution accuracy required more loops to improve the solution.

\begin{figure}
    \includegraphics[clip,width=1.0\linewidth]{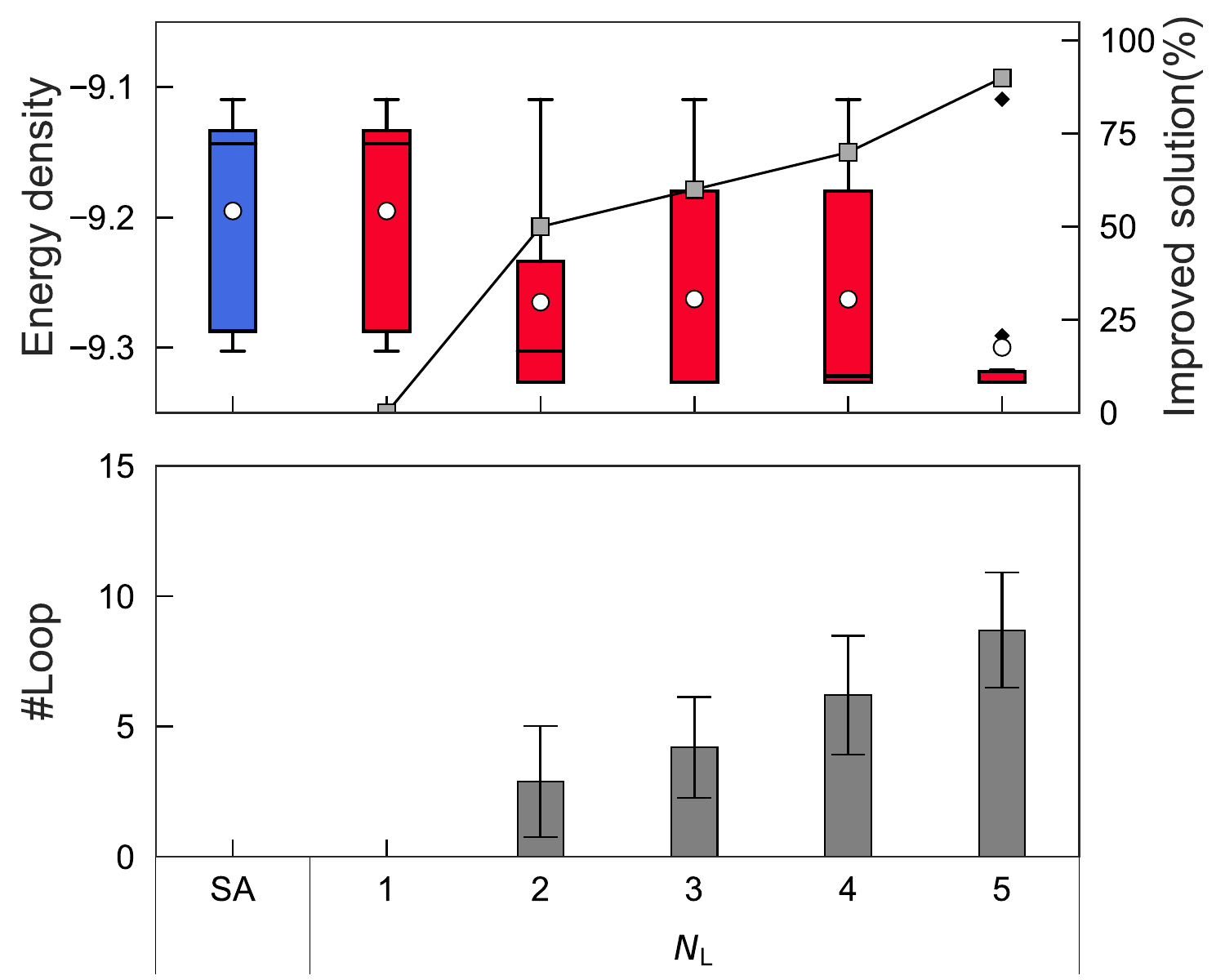}
    \caption{(Color online) Performance of hybrid optimization methods with different $N_{\rm L}$. (Top) Energy densities of preprocessing SA and the hybrid method. White circles are the average of ten runs. Top and bottom bars denote the highest and lowest energy densities, respectively. Black diamonds denote outliers. Gray squares denote the percentage of improved solutions from the preprocessing SA in the ten simulations of the hybrid optimization method. (Bottom) Number of loops by the hybrid optimization method. Error bars are standard deviations obtained in ten runs.}
    \label{fig:comparison_NL}
\end{figure}

Next, we performed the hybrid method with different parameter patterns to evaluate the effect of the parameters on the solution accuracy.
Figure~\ref{fig:comparison_NL} shows the energy density and the number of loops for $N_{\rm L}$ set from 1 to 5, while the outer loop was set to 50.
All other parameters were the same as those in Table~\ref{table:parameters_1}.
Note that the solution pool sets obtained by the preprocessing SA were for all conditions of $N_{\rm L}$.
In addition, Fig.~\ref{fig:comparison_NL} shows the improvement percentage in $X_{\textrm{best}}$ of the preprocessing SA among the ten simulations of the hybrid method.
The solutions were improved for $N_{\rm L} \ge 2$, although the solution with the highest energy corresponding to the top bar or an outlier in the box plot was not improved.

As $N_{\rm L}$ increased, both the percentage of improved solutions and the number of loops increased.
There are two possible reasons for the increased number of loops.
First, the hybrid method requires more loops prior to termination. 
Second, the solutions improve with an increase in $N_{\rm L}$. 

\begin{figure}
    \includegraphics[clip,width=1.0\linewidth]{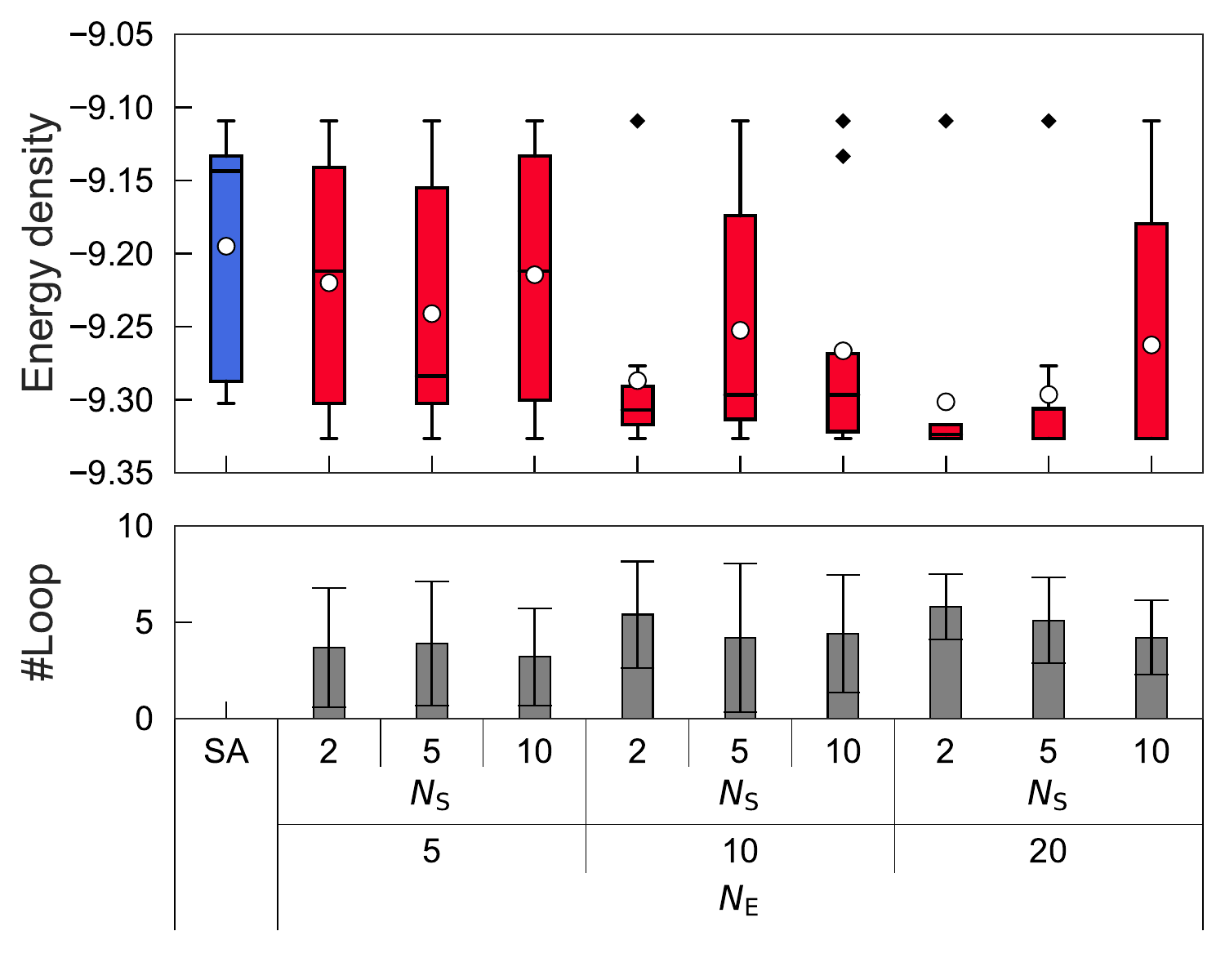}
    \caption{(Color online) Performance of hybrid optimization methods with different sets of $N_{\rm E}$ and $N_{\rm S}$. (Top) Energy densities of preprocessing SA and the hybrid method. White circles are the average of ten runs. Top and bottom bars denote the highest and lowest energy densities, respectively. Black diamonds denote outliers. (Bottom) Number of loops by the hybrid optimization method. Error bars are standard deviations.}
    \label{fig:comparison_NE_NS}
\end{figure}

Figure~\ref{fig:comparison_NE_NS} shows the energy density and the number of loops with some sets of $N_{\rm E}$ and $N_{\rm S}$.
Note that the sets of solution pools obtained by the preprocessing SA were the same for all conditions of $N_{\rm S}$ and $N_{\rm E}$, and the outer loop was set to 50.
All other parameters were the same as those in Table~\ref{table:parameters_1}.
The hybrid method improved the solution more when $N_{\rm E}$ was 10 or 20 compared to when $N_{\rm E}$ was 5.
This is because a larger number of $N_{\rm E}$ resulted in more solutions generated per loop, and there was a higher possibility of solution improvement prior to the loop termination.
Additionally, the effect of $N_{\rm S}$ was not significant when $N_{\rm E}$ was the same. 
The number of loops was higher for $N_{\rm E}$ of 10 and 20, which is where an improvement is observed.

\begin{figure}
    \includegraphics[clip,width=1.0\linewidth]{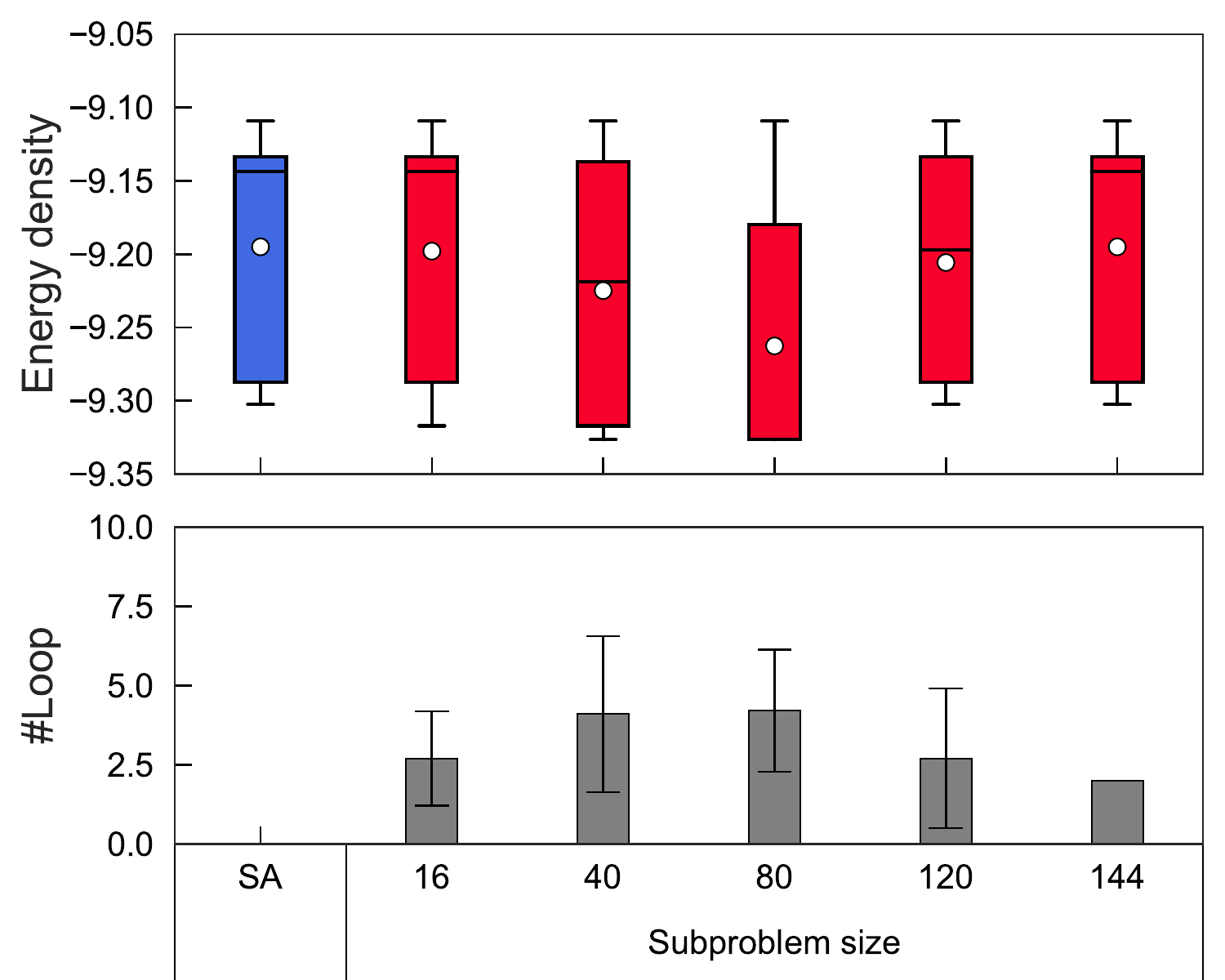}
    \caption{(Color online) Performance of hybrid optimization methods with different sub-Ising model size. (Top) Energy densities of preprocessing SA and the hybrid method. White circles are the average of ten runs. Top and bottom bars denote the highest and lowest energy densities, respectively. (Bottom) Number of loops by the hybrid method. Error bars are standard deviations.}
    \label{fig:comparison_m}
\end{figure}

Finally, we performed the hybrid method with different sub-Ising model sizes to evaluate the sub-Ising model size dependency.
Figure~\ref{fig:comparison_m} shows the energy density and the number of loops for the sub-Ising model sizes ($m$) of 16, 40, 80, 120, and 144.
Note that the sets of solution pools obtained by the preprocessing SA were for all sub-Ising model sizes.
The outer loop was set to 50.
All other parameters were the same as those in Table~\ref{table:parameters_1}.
Constructing sub-Ising models with extremely small or large numbers of spins (e.g., a sub-Ising model size $=$ 40 or 144) relative to the size of the original Ising model ($n=160$) did not improve the solution using the hybrid method (Fig.~\ref{fig:comparison_m}). 
When the sub-Ising model size was small, the fixed spins significantly influenced the sub-Ising model, and there was little freedom in the obtained solutions obtained.
By contrast, when the sub-Ising model size was large, the influence of the solution accuracy of D-Wave Advantage became more significant. 
Assuming that the solution accuracy of D-Wave Advantage for the original Ising model of size 160 was low (shown in Fig.~\ref{fig:comparison_Ising_model}), the sub-Ising model size became larger, the solution accuracy decreased, and the solution did not improve.
These results suggest that the solution accuracy of the hybrid method depends on the sub-Ising model size.

\section{Conclusion and future work}
\label{sec:conclusion}

We propose a hybrid optimization method, which utilizes the advantages of two types of Ising machines (non-quantum-type Ising machines and quantum annealers) 
Then we evaluated the performance of the hybrid method via simulations using SA as a non-quantum-type Ising machine and D-Wave Advantage as a quantum annealer.
The hybrid method achieves a higher solution accuracy than the D-Wave Advantage and preprocessing SA alone.
Regardless of the preprocessing accuracy, the hybrid method improves the solution.
However, there is a trade-off between the preprocessing accuracy and the number of loops for solution improvement. 

We also evaluated the effect of the hybrid method parameters on the solution accuracy.
Increasing the $N_{\rm L}$ value, which determines the termination condition of the hybrid method, improves the solution accuracy while simultaneously increasing the number of loops.
Next, we assessed the solution accuracy of the hybrid method with multiple combinations of $N_{\rm E}$ and $N_{\rm S}$.
The solution is enhanced when $N_{\rm E}$ is 10 or 20 compared to when $N_{\rm E}$ is 5.
However, the appropriate sub-Ising model spins should be selected because the hybrid method has a sub-Ising model dependency. 

This method first constructs a solution pool with a non-quantum-type Ising machine.
Then it solves $N_{\rm E}$ sub-Ising models using a quantum annealer. 
Using the hybrid method, it is more beneficial to solve multiple of the reduced-spin-size sub-Ising models embedding the original Ising model size into D-Wave Advantage.
If the sub-Ising models can obtaine solutions in parallel, the hybrid method should be accelerated using the concept of parallel quantum annealing~\cite{pelofske2022parallel}.
This approach should be considered in the future.

We also evaluated the performance of the hybrid method on the original Ising model with a problem size that can be embedded in D-Wave Advantage.
However, several methods~\cite{qbsolv, irie2021hybrid, atobe2021hybrid, pelofske2021decomposition} have been proposed to solve larger sizes of problems because the problem size that can be embedded in the quantum annealer is limited.
Therefore, the performance of the hybrid method should be evaluated in the future using an original Ising model that cannot be embedded in a quantum annealer.
Although the performance evaluation of the hybrid method was conducted using simulations with SA and D-Wave Advantage, the hybrid method should be evaluated using an actual Ising machine in the future.
Furthermore, a theoretical investigation on the origin of the improved solution accuracy by the proposed method is insufficient in this paper. 
Since a theoretical investigation should contribute to parameter setting and performance improvement, it remains as a future work. 

\section*{Acknowledgments}
This article is based on the results obtained from a project, JPNP16007, commissioned by the New Energy and Industrial Technology Development Organization (NEDO). 
The computations in this work were partially performed using the facilities of the Supercomputer Center, the Institute for Solid State Physics, The University of Tokyo.
S.~T. was supported in part by JSPS KAKENHI (Grant Numbers JP21K03391, JP23H05447) and JST Grant Number JPMJPF2221.
The Human Biology-Microbiome-Quantum Research Center (Bio2Q) is supported by the World Premier International Research Center Initiative (WPI), MEXT, Japan.    

\bibliography{reference.bib}

\begin{thebibliography}{10}

\bibitem{Johnson2011}
M.~W. Johnson, M.~H. Amin, S.~Gildert, T.~Lanting, F.~Hamze, N.~Dickson,
  R.~Harris, A.~J. Berkley, J.~Johansson, P.~Bunyk, E.~M. Chapple, C.~Enderud,
  J.~P. Hilton, K.~Karimi, E.~Ladizinsky, N.~Ladizinsky, T.~Oh, I.~Perminov,
  C.~Rich, M.~C. Thom, E.~Tolkacheva, C.~J.~S. Truncik, S.~Uchaikin, J.~Wang,
  B.~Wilson, and G.~Rose: Nature {\bfseries 473} (2011) 194.

\bibitem{Yamaoka2016}
M.~Yamaoka, C.~Yoshimura, M.~Hayashi, T.~Okuyama, H.~Aoki, and H.~Mizuno: IEEE
  J. Solid-State Circuits {\bfseries 51} (2016) 303.

\bibitem{Aramon2019}
M.~Aramon, G.~Rosenberg, E.~Valiante, T.~Miyazawa, H.~Tamura, and H.~G.
  Katzgraber: Front. Phys. {\bfseries 7} (2019) 48.

\bibitem{Inagaki2016}
T.~Inagaki, Y.~Haribara, K.~Igarashi, T.~Sonobe, S.~Tamate, T.~Honjo,
  A.~Marandi, P.~L. McMahon, T.~Umeki, K.~Enbutsu, O.~Tadanaga, H.~Takenouchi,
  K.~Aihara, K.-i. Kawarabayashi, K.~Inoue, S.~Utsunomiya, and H.~Takesue:
  Science {\bfseries 354} (2016) 603.

\bibitem{Goto2019}
H.~Goto, K.~Tatsumura, and A.~R. Dixon: Sci. Adv. {\bfseries 5} (2019)
  eaav2372.

\bibitem{Maezawa2019}
M.~{Maezawa}, G.~{Fujii}, M.~{Hidaka}, K.~{Imafuku}, K.~{Kikuchi}, H.~{Koike},
  K.~{Makise}, S.~{Nagasawa}, H.~{Nakagawa}, M.~{Ukibe}, and S.~{Kawabata}: J.
  Phys. Soc. Jpn. {\bfseries 88} (2019) 061012.

\bibitem{Motomura2020}
K.~{Yamamoto}, K.~{Kawamura}, K.~{Ando}, N.~{Mertig}, T.~{Takemoto},
  M.~{Yamaoka}, H.~{Teramoto}, A.~{Sakai}, S.~{Takamaeda-Yamazaki}, and
  M.~{Motomura}: IEEE J. Solid-State Circuits  (2020).

\bibitem{neven2009}
H.~Neven, V.~S. Denchev, M.~Drew-Brook, J.~Zhang, W.~G. Macready, and G.~Rose:
  Quantum {\bfseries 4} (2009).

\bibitem{Amin2015}
M.~H. Amin: Phys. Rev. A {\bfseries 92} (2015) 052323.

\bibitem{Amin2018}
M.~H. Amin, E.~Andriyash, J.~Rolfe, B.~Kulchytskyy, and R.~Melko: Phy. Rev. X
  {\bfseries 8} (2018) 021050.

\bibitem{Daniel2018}
D.~O'Malley, V.~V. Vesselinov, B.~S. Alexandrov, and L.~B. Alexandrov: PloS one
  {\bfseries 13} (2018) e0206653.

\bibitem{Kitai2020}
K.~Kitai, J.~Guo, S.~Ju, S.~Tanaka, K.~Tsuda, J.~Shiomi, and R.~Tamura: Phys.
  Rev. Res. {\bfseries 2} (2020) 013319.

\bibitem{utimula2021quantum}
K.~Utimula, T.~Ichibha, G.~I. Prayogo, K.~Hongo, K.~Nakano, and R.~Maezono:
  Sci. Rep. {\bfseries 11} (2021).

\bibitem{inoue2022towards}
T.~Inoue, Y.~Seki, S.~Tanaka, N.~Togawa, K.~Ishizaki, and S.~Noda: Opt. Express
  {\bfseries 30} (2022) 43503.

\bibitem{endo2022phase}
K.~Endo, Y.~Matsuda, S.~Tanaka, and M.~Muramatsu: Sci. Rep. {\bfseries 12}
  (2022) 10794.

\bibitem{sampei2023quantum}
H.~Sampei, K.~Saegusa, K.~Chishima, T.~Higo, S.~Tanaka, Y.~Yayama, M.~Nakamura,
  K.~Kimura, and Y.~Sekine: JACS Au {\bfseries 3} (2023) 991.

\bibitem{tucs2023quantum}
A.~Tu\v{c}s, F.~Berenger, A.~Yumoto, R.~Tamura, T.~Uzawa, and K.~Tsuda: ACS
  Med. Chem. Lett. {\bfseries 14} (2023) 577.

\bibitem{hatakeyama2023extracting}
K.~Hatakeyama-Sato, Y.~Uchima, T.~Kashikawa, K.~Kimura, and K.~Oyaizu: RSC Adv.
  {\bfseries 13} (2023) 14651.

\bibitem{rosenberg2016solving}
G.~Rosenberg, P.~Haghnegahdar, P.~Goddard, P.~Carr, K.~Wu, and M.~L. De~Prado:
  IEEE J. Sel. Top. Signal Processing {\bfseries 10} (2016) 1053.

\bibitem{Tanahashi2019}
K.~{Tanahashi}, S.~{Takayanagi}, T.~{Motohashi}, and S.~{Tanaka}: J. Phys. Soc.
  Jpn. {\bfseries 88} (2019) 061010.

\bibitem{Perdomo-Ortiz2012}
A.~Perdomo-Ortiz, N.~Dickson, M.~Drew-Brook, G.~Rose, and A.~Aspuru-Guzik: Sci.
  Rep. {\bfseries 2} (2012).

\bibitem{Neukart2017}
F.~Neukart, G.~Compostella, C.~Seidel, D.~von Dollen, S.~Yarkoni, and
  B.~Parney: Front. ICT {\bfseries 4} (2017) 29.

\bibitem{irie2019quantum}
H.~Irie, G.~Wongpaisarnsin, M.~Terabe, A.~Miki, and S.~Taguchi: International
  Workshop on Quantum Technology and Optimization Problems, 2019, p. 145.

\bibitem{Bao2021-a}
S.~Bao, M.~Tawada, S.~Tanaka, and N.~Togawa: 2021 International Symposium on
  VLSI Design, Automation and Test (VLSI-DAT), 2021.

\bibitem{Bao2021-b}
S.~Bao, M.~Tawada, S.~Tanaka, and N.~Togawa: 2021 IEEE International
  Intelligent Transportation Systems Conference (ITSC), 2021, p. 3704.

\bibitem{Mukasa2021}
Y.~Mukasa, T.~Wakaizumi, S.~Tanaka, and N.~Togawa: IEICE TRANSACTIONS on
  Information and Systems {\bfseries 104} (2021) 1592.

\bibitem{naito2023isaaq}
S.~Naito, Y.~Hasegawa, Y.~Matsuda, and S.~Tanaka: arXiv:2303.02830 .

\bibitem{izawa2022continuous}
S.~Izawa, K.~Kitai, S.~Tanaka, R.~Tamura, and K.~Tsuda: Phys. Rev. Res.
  {\bfseries 4} (2022) 023062.

\bibitem{seki2022black}
Y.~Seki, R.~Tamura, and S.~Tanaka: arXiv:2209.01016 .

\bibitem{kochenberger2014unconstrained}
G.~Kochenberger, J.-K. Hao, F.~Glover, M.~Lewis, Z.~L{\"u}, H.~Wang, and
  Y.~Wang: J. Comb. Optim. {\bfseries 28} (2014) 58.

\bibitem{lucas2014ising}
A.~Lucas: Front. Phys. {\bfseries 2} (2014).

\bibitem{Tanaka2017}
S.~Tanaka, R.~Tamura, and B.~K. Chakrabarti: {\em Quantum Spin Glasses,
  Annealing and Computation} (Cambridge University Press, 2017).

\bibitem{Tanaka2020}
S.~{Tanaka}, Y.~{Matsuda}, and N.~{Togawa}: 2020 25th Asia and South Pacific
  Design Automation Conference (ASP-DAC), 2020, p. 659.

\bibitem{mohseni2022ising}
N.~Mohseni, P.~L. McMahon, and T.~Byrnes: Nature Reviews Physics {\bfseries 4}
  (2022) 363.

\bibitem{kadowaki1998quantum}
T.~Kadowaki and H.~Nishimori: Phys. Rev. E {\bfseries 58} (1998) 5355.

\bibitem{kirkpatrick1983optimization}
S.~Kirkpatrick, C.~D. Gelatt, and M.~P. Vecchi: Science {\bfseries 220} (1983)
  671.

\bibitem{johnson1989optimization}
D.~S. Johnson, C.~R. Aragon, L.~A. McGeoch, and C.~Schevon: Oper. Res.
  {\bfseries 37} (1989) 865.

\bibitem{johnson1991optimization}
D.~S. Johnson, C.~R. Aragon, L.~A. McGeoch, and C.~Schevon: Oper. Res.
  {\bfseries 39} (1991) 378.

\bibitem{Oku2020}
D.~Oku, M.~Tawada, S.~Tanaka, and N.~Togawa: IEEE Trans. Comput. {\bfseries 71}
  (2022) 223.

\bibitem{Kowalsky2021}
M.~Kowalsky, T.~Albash, I.~Hen, and D.~A. Lidar: Quantum Sci. Technol.
  {\bfseries 7} (2022) 025008.

\bibitem{rosenberg2016building}
G.~Rosenberg, M.~Vazifeh, B.~Woods, and E.~Haber: Comput. Optim. Appl.
  {\bfseries 65} (2016) 845.

\bibitem{qbsolv}
M.~{Booth}, S.~P. {Reinhardt}, and A.~{Roy}: D-Wave Technical Report Series
  14-1006A-A, 2017.
\newblock
  [https://docs.ocean.dwavesys.com/projects/qbsolv/en/latest/index.html].

\bibitem{karimi2017effective}
H.~Karimi, G.~Rosenberg, and H.~G. Katzgraber: Phys. Rev. E {\bfseries 96}
  (2017) 043312.

\bibitem{karimi2017boosting}
H.~Karimi and G.~Rosenberg: Quantum Inf. Process {\bfseries 16} (2017) 166.

\bibitem{irie2021hybrid}
H.~Irie, H.~Liang, T.~Doi, S.~Gongyo, and T.~Hatsuda: Sci. Rep. {\bfseries 11}
  (2021) 8426.

\bibitem{atobe2021hybrid}
Y.~Atobe, M.~Tawada, and N.~Togawa: IEEE Trans. Comput. {\bfseries 71} (2022)
  2606.

\bibitem{pelofske2021decomposition}
E.~Pelofske, G.~Hahn, and H.~Djidjev: Journal of Signal Processing Systems
  {\bfseries 93} (2021) 405.

\bibitem{pelofske2023solving}
E.~Pelofske, G.~Hahn, and H.~N. Djidjev: Quantum Inf. Process. {\bfseries 22}
  (2023) 219.

\bibitem{Yoshimura2017}
C.~Yoshimura, M.~Hayashi, T.~Okuyama, and M.~Yamaoka: International Journal of
  Networking and Computing {\bfseries 7} (2017) 154.

\bibitem{Okuyama2017}
T.~Okuyama, M.~Hayashi, and M.~Yamaoka: 2017 IEEE International Conference on
  Rebooting Computing, ICRC 2017 - Proceedings {\bfseries 2017-Janua} (2017).

\bibitem{yamamoto2020}
K.~Yamamoto, K.~Ando, N.~Mertig, T.~Takemoto, M.~Yamaoka, H.~Teramoto,
  A.~Sakai, S.~Takamaeda-Yamazaki, and M.~Motomura: 2020 IEEE International
  Solid-State Circuits Conference-(ISSCC), 2020, p. 138.

\bibitem{FixAE}
{Fixstars Amplify Annealing Engine: Fixstars Amplify}.
\newblock \url{[https://amplify.fixstars.com/en/}].

\bibitem{Ocean}
{Solver Parameters - D-Wave System Documentation}.
\newblock
  \url{[https://docs.dwavesys.com/docs/latest/c_solver_parameters.html]}.

\bibitem{dattani2019pegasus}
N.~Dattani, S.~Szalay, and N.~Chancellor: arXiv:1901.07636 .

\bibitem{DW_update}
{The Advantage System: Performance Update}.
\newblock
  \url{[https://www.dwavesys.com/media/kjtlcemb/14-1054a-a_advantage_system_performance_update.pdf]}.

\bibitem{pelofske2022parallel}
E.~Pelofske, G.~Hahn, and H.~N. Djidjev: Sci. Rep. {\bfseries 12} (2022).

\end{thebibliography}

\end{document}